\documentclass[10pt,aps,prd,twocolumn,fleqn,floatfix,nofootinbib]{revtex4-2}

\usepackage[section]{placeins}
\usepackage{amsmath,amssymb,amsfonts}
\usepackage{graphicx}
\usepackage[usenames,dvipsnames]{xcolor}
\usepackage{mathtools}
\usepackage{bm}
\usepackage{microtype}
\usepackage[breaklinks=true,colorlinks=true,
            linkcolor=MidnightBlue,citecolor=MidnightBlue,urlcolor=MidnightBlue]{hyperref}

\allowdisplaybreaks[1]
\graphicspath{{figures/}}
\newcommand{\SYM}{\mathcal N=4~\mathrm{SYM}}
\newcommand{\Lam}{\lambda}

\newcommand{\Alog}{A_{2\log}}

\begin{document}

\title{Constrained Pad\'e Ensembles for Thermal $\mathcal{N}{=}4$ SYM
with the Exact $\mathcal O(\lambda^{5/2})$ Coefficient}

\author{Ubaid Tantary}
\email{utantary@pmu.edu.sa}
\affiliation{Department of Mathematics and Natural Sciences,
Prince Mohammad Bin Fahd University, Al Khobar 31952, Saudi Arabia}

\author{Qianqian Du}
\email{dqianqian@mailbox.gxnu.edu.cn}
\affiliation{Department of Physics, Guangxi Normal University,
Guilin 541004, China}
\affiliation{Guangxi Key Laboratory of Nuclear Physics and Technology,
Guilin 541004, China}

\date{\today}

\begin{abstract}
We revisit the constrained log-subtracted two-point Pad\'e (LSTP) ensemble
for thermal $\mathcal{N}{=}4$ supersymmetric Yang-Mills (SYM)
thermodynamics in four spacetime dimensions after upgrading the weak-coupling
truncation from $\mathcal O(\Lam^2)$ to the exact $\mathcal O(\Lam^{5/2})$
coefficient. We keep the LSTP interpolation ansatz unchanged and shift the
weak-side matching points to the regime where the new term is
numerically significant. Within the same parameter scan the admissible set
collapses from $9$ nominal survivors
($3$ distinct curves) to a single distinct curve, the crossover range
shrinks to a unique value,
and the pointwise band width drops to zero within numerical resolution.
The Hermite-Pad\'e (HP) central
curve does not coincide with the unique LSTP
survivor, so the updated weak-side matching removes the LSTP scan
uncertainty but not the difference between the two routes. The minimal
HP extension that also carries the exact coefficient is uniquely
determined and develops a spurious pole-zero pair at small coupling.
Away from that pair the HP and LSTP curves still differ, so this
difference persists at the same weak-coupling order.
\end{abstract}

\maketitle

\section{Introduction}

The thermodynamics of thermal $\SYM$ is described by a single function
$f(\Lam)\equiv\mathcal S/\mathcal S_0$, the Stefan--Boltzmann-normalized
entropy ratio. Perturbation theory fixes the
small-$\Lam$ expansion~\cite{Du:2021jai,Andersen:2021kld,Carrington:2026},
while AdS/CFT fixes the large-$\Lam$
limit~\cite{Maldacena:1997re,Gubser:1998bc}. Neither expansion converges
in the crossover region, so intermediate coupling can only be reached
by interpolation (see also Ref.~\cite{Muller:2025}). In
Ref.~\cite{Tantary:2026ensemble} we explored this problem with
constrained Pad\'e ensembles matched to the weak-coupling expansion
through $\mathcal O(\Lam^2)$. The main result was not a single preferred
interpolant but an admissible family with an uncertainty band. The
recent calculation of the exact $\mathcal O(\Lam^{5/2})$
coefficient~\cite{Carrington:2026} lets us test whether a
higher-order weak-coupling coefficient narrows the ensemble or selects
a unique survivor.

The new coefficient provides an immediate check: the Hermite--Pad\'e (HP) ansatz used in 
Ref.~\cite{Tantary:2026ensemble} predicted
$A_{5/2}^{\mathrm{HP}}=-43.8\pm0.1$ (in the normalization of
Eq.~\eqref{eq:weak52} below), whereas the exact result is
$A_{5/2}\simeq+0.754$, differing in both sign and magnitude. This
discrepancy is consistent with the spurious $\Lam^{5/2}\log\Lam$
artifact already noted in Ref.~\cite{Tantary:2026ensemble}, which
means the HP rational structure is unreliable at unmatched
orders. Rebuilding the HP route requires extending the ansatz itself;
the minimal such extension is constructed in Sec.~\ref{sec:hp_update}
and is uniquely determined. It develops a near-canceling spurious
pole-zero pair, which we identify and classify there; evaluated away
from the pair, the curve remains distinct from the updated LSTP
survivor. We focus on the updated LSTP scan, keeping the $[5/5]$ HP
curve unchanged so we can compare routes.

The band does not merely narrow. We revisit the same
log-subtracted two-point Pad\'e (LSTP) scan used in
Ref.~\cite{Tantary:2026ensemble}, keeping the rational order, mapping
family, admissibility filters, and numerical strategy fixed. We then
replace the weak-side truncation with the exact $\mathcal O(\Lam^{5/2})$
expansion and shift the weak-side matching points to intermediate
couplings where the new term is numerically significant. The earlier
$\mathcal O(\Lam^2)$ scan had $9$ nominal survivors, $3$ distinct
curves, and a broad crossover interval. With the exact $\mathcal
O(\Lam^{5/2})$ expansion, only $3$ nominal survivors remain, all sharing
the log-subtraction cutoff scale $\Lambda_0=4$ (defined in Sec.~\ref{sec:method}),
and they are numerically identical. In other words, the
admissible LSTP band collapses to a unique curve.

The collapse is one finding; a second follows at once. The HP
central curve retained unchanged from the earlier
analysis does \emph{not} coincide with the unique LSTP survivor. The
updated scan therefore removes the ambiguity within the
LSTP family but does not close the gap between the LSTP curve and the
HP curve. It pins down a unique LSTP curve but does
not resolve the LSTP--HP disagreement.
\section{Weak- and strong-coupling expansions}
\label{sec:inputs}

We use the 't~Hooft coupling $\Lam=g^2N_c$. At weak coupling the
Stefan--Boltzmann-normalized entropy ratio is
\begin{equation}
\label{eq:weak52}
\begin{aligned}
f(\Lam)
&= 1 - \frac{3}{2\pi^2}\,\Lam
   + \frac{3+\sqrt{2}}{\pi^3}\,\Lam^{3/2}
   + A_{2\log}\,\Lam^2\log\!\left(\frac{\Lam}{\pi^2}\right) \\
&\quad + A_{20}\left(\frac{\Lam}{\pi^2}\right)^{\!2}
   + A_{5/2}\,\frac{\Lam^{5/2}}{\pi^5}
   + \mathcal O(\Lam^3\log\Lam)\,.
\end{aligned}
\end{equation}
The coefficients $A_{2\log}$ and $A_{20}$ were calculated using resummation and effective field theory (EFT) methods in
Refs.~\cite{Du:2021jai,Andersen:2021kld},
\begin{align}
A_{2\log} &= \frac{3}{2\pi^4}\,,\\[2pt]
A_{20}
&= -\frac{21}{8} - \frac{9\sqrt{2}}{8}
  + \frac{3}{2}\,\gamma_E
  + \frac{3}{2}\,\frac{\zeta'(-1)}{\zeta(-1)}
  - \frac{25\log 2}{8}\,,
\end{align}
where $\gamma_E$ is the Euler--Mascheroni constant. The exact
$\mathcal O(\Lam^{5/2})$ coefficient was obtained in
Ref.~\cite{Carrington:2026},
\begin{equation}
\label{eq:A52}
\begin{aligned}
A_{5/2}
&=\frac{33}{8}\bigl(\log 4-1\bigr)
+\frac{3}{128}(28+25\sqrt{2})\log(1+\sqrt{2}) \\
&\quad -\frac{1}{64}(6+\sqrt{2})\pi^2
-\frac{35+212\log 2}{128\sqrt{2}} \\
&\simeq 0.7537170300\,.
\end{aligned}
\end{equation}
The $\mathcal O(\Lam^{5/2})$ calculation does not contribute at order
$\Lam^2\log\Lam$, so $A_{2\log}$ is carried forward unchanged.
Throughout this paper we run two parallel LSTP scans with identical
construction: a \emph{before} scan that truncates the weak series at
$\mathcal O(\Lam^2)$ (matching Ref.~\cite{Tantary:2026ensemble}, i.e.,
with no $A_{5/2}$) and an \emph{after} scan that includes the exact
$A_{5/2}$ and extends the weak truncation to $\mathcal O(\Lam^{5/2})$.
The two scans share the same ansatz, conformal map, admissibility
filters, $(\alpha,\Lambda_0)$ grid, and strong-side matching points.
The weak-side content differs in two linked ways: the \emph{after}
scan adds $A_{5/2}$ to the weak target and shifts the weak-side
matching points to intermediate couplings where $A_{5/2}$ is
numerically significant (Sec.~\ref{sec:method}). The two changes are
introduced together, so any change in the admissible set between the
two scans reflects their combined effect.

At strong coupling (large $N_c$)~\cite{Gubser:1998bc,Maldacena:1997re},
\begin{equation}
\label{eq:strong}
f(\Lam)=\frac{3}{4}\!\left[1+\frac{15}{8}\zeta(3)\,\Lam^{-3/2}
+\mathcal O(\Lam^{-3})\right],
\end{equation}
with no $\Lam^{-1/2}$ or $\Lam^{-1}$ terms.
\section{Updated LSTP construction}
\label{sec:method}

We keep the interpolation framework of Ref.~\cite{Tantary:2026ensemble}
intact and ask what changes when the exact weak-side coefficient $A_{5/2}$
is imposed. Specifically, we keep the log-subtracted Pad\'e ansatz
\begin{equation}
\label{eq:gdef}
g(\Lam)=f(\Lam)-\Alog\,\Lam^2\log\!\left(\frac{\Lam}{\pi^2}\right)
\,\chi(\Lam;\Lambda_0,p)\,,
\end{equation}
with $\Alog=3/(2\pi^4)$ and
\begin{equation}
\chi(\Lam;\Lambda_0,p)=\frac{1}{1+(\Lam/\Lambda_0)^p}\,,
\qquad p=3\,.
\end{equation}
We keep $p=3$ as in Ref.~\cite{Tantary:2026ensemble}; this is the
minimal integer for which the cutoff-induced leakage from the
log-subtraction term decays at large $\Lam$. Using the same $p$
ensures that any change in the admissible set comes from the updated
weak-side matching, not from the cutoff.

We approximate the log-subtracted function $g(\Lam)$ by a rational
function $R(z)$ in the conformal variable
\begin{equation}
z=\frac{\sqrt{\Lam}}{1+\alpha\sqrt{\Lam}}\,,
\end{equation}
which corresponds to setting $\beta=0$ in the two-parameter M\"obius family
$z=\sqrt{\Lam}/(1+\alpha\sqrt{\Lam}+\beta\Lam)$ used in
Ref.~\cite{Tantary:2026ensemble}. We restrict to $\beta=0$ because this sub-family already contains
every admissible LSTP survivor reported in
Ref.~\cite{Tantary:2026ensemble}; a single mapping parameter $\alpha$
is therefore sufficient. That is, we set
\begin{equation}
g(\Lam)\;\approx\;R\bigl(z(\Lam)\bigr),
\end{equation}
where $R(z)$ is a near-diagonal $[4/4]$ Pad\'e,
\begin{equation}
\label{eq:pade44}
R(z)=\frac{a_0+a_1 z+a_2 z^2+a_3 z^3+a_4 z^4}
           {1+b_1 z+b_2 z^2+b_3 z^3+b_4 z^4}\,.
\end{equation}
Here $R(z)$ is not a ratio $g/f$: it is a rational approximant
to $g(\Lam)$ written in the $z$ variable, and $f(\Lam)$ is recovered
by inverting Eq.~\eqref{eq:gdef}, $f(\Lam)=R(z(\Lam))
+\Alog\Lam^2\log(\Lam/\pi^2)\,\chi(\Lam;\Lambda_0,p)$.
Equation~\eqref{eq:pade44} has $5+4=9$ unknown coefficients (five in
the numerator and four in the denominator, the constant term of the
denominator fixed to unity). We use the same bounds and monotonicity
filters as before, together with the mapped-root test:
\begin{enumerate}
\item $0.75\le f(\Lam)\le1$ on the evaluation grid,
\item $df/d(\log\Lam)\le 0$ on the evaluation grid,
\item no isolated denominator roots of the mapped rational approximant
that map into the numerical positive-axis window
$\operatorname{Re}\Lam>10^{-6}$ and
$|\operatorname{Im}\Lam|<0.05|\operatorname{Re}\Lam|$, including roots
outside the physical interval $0<z<1/\alpha$.
\end{enumerate}
Filter 3 is a criterion on the roots of the mapped rational
approximant, not a claim that every excluded root represents a physical
singularity. In Sec.~\ref{sec:hp_update} it is refined to
distinguish isolated denominator roots from near-canceling spurious
pole-zero pairs; the refinement leaves the scan reported below unchanged.
The $9$ unknown coefficients in Eq.~\eqref{eq:pade44} are fixed by
imposing $9$ matching conditions on $R(z(\Lam))$: we require
$R(z(\Lam_i))$ to equal the truncated target values at five
weak-side points $\Lam_i$ and at four strong-side points $\Lam_j$
(the specific values of $\Lam_i,\Lam_j$ are given below). The ``9'' and the ``5'' refer to the number of matching
\emph{points}, not to the number of Taylor coefficients; they are set
by the $[4/4]$ rational order, not by the orders of the truncated
expansions. All small-$\Lam$ contributions in
Eq.~\eqref{eq:weak52} (the Stefan--Boltzmann limit $f(0)=1$, the
coefficients $A_1=-\tfrac{3}{2\pi^2}$ and
$A_{3/2}=\tfrac{3+\sqrt{2}}{\pi^3}$ at orders $\Lam$ and $\Lam^{3/2}$,
$A_{20}$, $A_{5/2}$, and the logarithmic coefficient $A_{2\log}$) are
retained in the construction. The
Stefan--Boltzmann constant and $A_1,A_{3/2},A_{20},A_{5/2}$ all enter
through the numerical values of $f_{\rm weak}(\Lam_i)$ at the weak-side
matching points, where the series $1+A_1\Lam+\cdots$ is
evaluated. The logarithmic coefficient $A_{2\log}$ enters through the
log subtraction in Eq.~\eqref{eq:gdef}, which removes the $\Lam^2\log\Lam$
term from $g(\Lam)$ so that a purely rational ansatz for $g$ is
consistent. We do not place a matching point exactly at $\Lam=0$. The smallest
weak-side point ($\Lam=10^{-6}$ for the $\mathcal O(\Lam^2)$ scan,
$\Lam=10^{-4}$ for the $\mathcal O(\Lam^{5/2})$ scan) already sets
$f_{\rm weak}(\Lam_{\min})$ to unity with a deviation of
$\mathcal O(\Lam_{\min})$ (about $10^{-7}$ and $10^{-5}$
respectively), which effectively fixes $a_0\simeq 1$. Adding $\Lam=0$
as a separate matching condition would therefore be redundant. This yields a linear system of $9$
equations in $9$ unknowns with a unique solution per
$(\alpha,\Lambda_0)$.
Both the $\mathcal O(\Lam^2)$ and $\mathcal O(\Lam^{5/2})$
scans reported below use this same construction.

To compare directly with Ref.~\cite{Tantary:2026ensemble} we use the
same $\alpha$--$\Lambda_0$ grid,
\begin{equation}
\alpha\in\{0.5,1.0,2.0\},
\qquad
\Lambda_0\in\{0.5,1.0,2.0,4.0\}\,.
\end{equation}
The $\mathcal O(\Lam^2)$ scan uses weak-side matching points
$\Lam\in\{10^{-6},10^{-5},10^{-4},10^{-3},10^{-2}\}$, while the updated
$\mathcal O(\Lam^{5/2})$ scan shifts these to intermediate couplings
$\Lam\in\{10^{-4},10^{-3},0.1,0.5,1.0\}$ where the new term is
numerically significant. Both scans share the same strong-side matching
points $\Lam\in\{10,30,100,300\}$.

Within the $\beta=0$ subfamily, the three values of $\alpha$ give
numerically identical curves at each surviving $\Lambda_0$. This
$\alpha$-degeneracy is a numerical consequence of the specific
$9\times 9$ system built with these parameter choices; it is not an
underlying symmetry of the construction.

\section{Hermite--Pad\'e interpolation}
\label{sec:hp_ref}

\subsection{Reference curve}
For comparison we also use the Hermite--Pad\'e (HP) interpolant of
Ref.~\cite{Tantary:2026ensemble}, a generalized two-point $[5/5]$
Pad\'e in $y=\sqrt{\Lam}$, in the form introduced in
Refs.~\cite{Du:2021jai,Andersen:2021kld} (the HP route works in
$\sqrt{\Lam}$ directly, not in the LSTP variable
$z=\sqrt{\Lam}/(1+\alpha\sqrt{\Lam})$),
\begin{equation}
\label{eq:hp_ansatz}
f(\Lam)=\frac{1+a\,y+b(\Lam)\,y^2+c\,y^3+d\,y^4+e(\Lam)\,y^5}
             {1+a\,y+\bar b(\Lam)\,y^2+\tfrac43 c\,y^3+\tfrac43 d\,y^4
              +\tfrac43 e(\Lam)\,y^5},
\end{equation}
with $b(\Lam)=b_0+\pi^{-2}\log(\Lam/\pi^2)$ carrying the
$\Lam^2\log\Lam$ term of Eq.~\eqref{eq:weak52} and
$\bar b(\Lam)=b(\Lam)+3/(2\pi^2)$, the shift that reproduces the
$\mathcal O(\Lam)$ term. The strong-coupling matching fixes the fifth-order coefficient
$e(\Lam)$ in terms of $b(\Lam)$,
\begin{equation}
e(\Lam)=\frac{2b(\Lam)}{15\zeta(3)}-\frac{3}{5\pi^2\zeta(3)}.
\end{equation}
The logarithm carried by $b(\Lam)$ then reappears in $e(\Lam)$, rescaled
by $2/[15\zeta(3)]$, so $e(\Lam)$ introduces no independent logarithmic
coefficient. The factor $4/3$ in the denominator enforces $f\to 3/4$ while
removing the $\Lam^{-1/2}$ and $\Lam^{-1}$ strong-coupling terms. The
constants $a,b_0,c,d$ are fixed by matching the weak expansion through
$\mathcal O(\Lam^2)$ and the strong expansion through
$\mathcal O(\Lam^{-3/2})$, with no free parameter:
\begin{align}
a &= \frac{4\pi^2}{135\,\zeta(3)}+\frac{2(3+\sqrt2)}{3\pi}\approx 1.1800,\\
c &= \frac{2}{15\,\zeta(3)}\approx 0.1109,\\
d &= \frac{180(3+\sqrt2)\,\zeta(3)+8\pi^3}{2025\,\pi\,\zeta(3)^2}\approx 0.1309,\\
b_0 &= \frac{16\pi\bigl(45(3+\sqrt2)\,\zeta(3)+\pi^3\bigr)}
            {18225\,\zeta(3)^2}\notag\\
&\quad+\frac{36\bigl(\zeta'(-1)/\zeta(-1)+\gamma_E\bigr)
              +69\sqrt2+59-75\log 2}{36\pi^2}\notag\\
&\quad\approx 1.0689.
\end{align} We retain this HP curve unchanged, as a reference for
comparing routes.

\subsection{Symmetric update with the exact $A_{5/2}$}
\label{sec:hp_update}

The updated LSTP curve does not coincide with the HP reference
(Sec.~\ref{sec:results}). To compare the two constructions at the same
weak-coupling order, we now impose the exact $A_{5/2}$ in the HP
construction. The existing HP ansatz is already fixed by matching through
$\mathcal O(\Lam^2)$ at weak coupling and $\mathcal O(\Lam^{-3/2})$ at
strong coupling, with no free parameter, so $A_{5/2}$ cannot be imposed
without extending it. The $[5/5]$ rational structure also generates a
$\Lam^{5/2}\log\Lam$ term at its first unmatched weak-coupling order,
which Eq.~\eqref{eq:weak52} does not contain~\cite{Tantary:2026ensemble};
an extension must therefore both incorporate $A_{5/2}$ and cancel this
spurious logarithm. To obtain a nonzero finite strong-coupling limit $f\to\tfrac34$
within this rational family, the numerator and denominator must have the
same degree. The minimal equal-degree extension of this logarithmic
ansatz is the two-point $[6/6]$ Pad\'e in $y=\sqrt{\Lam}$,
\begin{equation}
\label{eq:hp66}
f_{\rm HP66}(\Lam)=\frac{N_6(y,L)}{D_6(y,L)},
\qquad y=\sqrt{\Lam},\quad L=\log(\Lam/\pi^2),
\end{equation}
with
\begin{align}
\label{eq:hp66ND}
N_6 &=1+n_1y+(n_{20}+n_{2L}L)\,y^2+n_3y^3+n_4y^4\notag\\
    &\quad +(n_{50}+n_{5L}L)\,y^5+n_6y^6,\\
D_6 &=1+d_1y+(d_{20}+d_{2L}L)\,y^2+d_3y^3+d_4y^4\notag\\
    &\quad +(d_{50}+d_{5L}L)\,y^5+d_6y^6,\notag
\end{align}
the logarithms entering through the $y^2$ and $y^5$ coefficients. Those
at $y^2$ provide the freedom needed to cancel unwanted logarithms below
order $\Lam^2$ while reproducing the required $\Lam^2\log\Lam$ term of
Eq.~\eqref{eq:weak52}, as in the $[5/5]$ curve; those at $y^5$ provide
the freedom to cancel the $\Lam^{5/2}\log\Lam$ term that the rational
form would otherwise produce. The coefficients are fixed by
\begin{itemize}
\item the weak-coupling coefficients of Eq.~\eqref{eq:weak52} through
$\mathcal O(\Lam^{5/2})$;
\item the absence of a $\Lam^{5/2}\log\Lam$ term;
\item the strong limit $f\to\tfrac34$, i.e.\ $d_6=\tfrac43 n_6$;
\item the absence of the $\Lam^{-1/2}$, $\Lam^{-1}$, $\Lam^{-2}$, and
$\Lam^{-5/2}$ terms, as in the strong expansion of
Eq.~\eqref{eq:strong};
\item the known $\Lam^{-3/2}$ coefficient $\tfrac{15}{8}\zeta(3)$ of
Eq.~\eqref{eq:strong}.
\end{itemize}
Equation~\eqref{eq:hp66ND} has $8+7=15$ unknown coefficients (eight in
the numerator $N_6$ and seven in the denominator $D_6$, the last
denominator coefficient fixed to $d_6=\tfrac43 n_6$ by the strong
limit). The eight weak-side conditions are the absence of the
$\Lam^{1/2}$ term (i.e.\ $n_1=d_1$), the constant and logarithmic parts
at orders $\Lam$, $\Lam^2$, and $\Lam^{5/2}$, and the known
$\Lam^{3/2}$ coefficient. The seven strong-side conditions are the
constant and logarithmic parts at orders $\Lam^{-1/2}$ and $\Lam^{-2}$,
the absence of the $\Lam^{-1}$ and $\Lam^{-5/2}$ terms, and the known
$\Lam^{-3/2}$ coefficient. The matching is therefore a linear system of
$15$ equations for $15$ independent coefficients with a unique solution.
The solution reproduces every term displayed in Eq.~\eqref{eq:weak52}
and generates no $\Lam^{5/2}\log\Lam$ term. At the first unmatched
order the rational form generates terms in
$\Lam^3[\log(\Lam/\pi^2)]^2$, unlike the
$\mathcal O(\Lam^3\log\Lam)$ remainder of Eq.~\eqref{eq:weak52}. No
result reported below uses the expansion at this order.

The unique solution has a near-canceling pole-zero pair on the
positive real $\Lam$ axis: $D_6$ vanishes at
$\Lam_D=0.179556755567488\ldots$ and $N_6$ at
$\Lam_N=0.179556946766974\ldots$, a separation of
$1.9\times10^{-7}$ that is not resolved on the evaluation
grid, where the curve appears smooth. Normalized to
$\Lam_D$, the separation and the residue of $f_{\rm HP66}$ at the
pole are
\begin{align}
\label{eq:doublet}
&\frac{|\Lam_N-\Lam_D|}{\Lam_D}=1.06\times10^{-6},\notag\\
&\frac{\bigl|{\rm Res}_{\Lam_D} f_{\rm HP66}\bigr|}{\Lam_D}
=1.04\times10^{-6}.
\end{align}
A pole with such a small residue, paired with a nearly coincident zero,
is a Froissart
doublet~\cite{Gonnet:2013}: the numerator and denominator of
Eq.~\eqref{eq:hp66ND} share an almost common factor, so away from the
pair the curve behaves as a rational function of lower order. We treat
it as a localized artifact of the rational approximation, not a
physical singularity of $f(\Lam)$.

The doublet does not arise from rounding error. For the matching equations as
written, the $15\times15$ coefficient matrix has full numerical rank,
with singular values between $67.3$ and $3.57\times10^{-3}$, and the
pair survives in an $80$-digit solution. Motivated by
Ref.~\cite{Gonnet:2013}, removing the smallest singular value
eliminates the pair, but the resulting rank-$14$ solution
violates the two exact logarithmic conditions by
$-9.8\times10^{-3}$ and $-8.3\times10^{-3}$, the first being $64\%$ of
$A_{2\log}=3/(2\pi^4)\simeq0.0154$. Since this solution no longer
reproduces the known weak-coupling logarithmic structure, it is not
considered further.

Dividing out the almost common factor,
$f_{\rm HP66}\to f_{\rm HP66}\,(\Lam-\Lam_D)/(\Lam-\Lam_N)$, removes
the pole and leaves the curve bounded and monotone. At $\Lam=2.9825$
and $20$, the relative differences from the exactly matched curve are
$7\times10^{-8}$ and $1\times10^{-8}$, respectively; the crossover is
unchanged to four digits, and $\hat S_3(20)$ changes by less than
$10^{-4}$. Because this shifts the Stefan--Boltzmann limit by
$1.1\times10^{-6}$ and introduces a $\Lam^{-1}$ term of order
$10^{-7}$ at strong coupling, it serves only as a stability check;
all values quoted below come from the exactly matched curve.

Evaluated away from the pair, the crossover and strong-coupling
residual of the exactly matched $[6/6]$ curve are
\begin{align}
\label{eq:hp66results}
&\Lam_c^{\rm HP66}\simeq 2.9825,\qquad
f(\Lam_c^{\rm HP66})\simeq 0.8588,\notag\\
&\hat S_3^{\rm HP66}(20)\simeq-29.77,
\end{align}
to be compared with $\Lam_c^{\rm LSTP}\simeq4.79$ and
$\hat S_3^{\rm LSTP}(20)\simeq+2.14$ for the updated LSTP curve
(Sec.~\ref{sec:results}). Within the two constructions built from the
same known weak- and strong-coupling expansions, the HP estimates
remain distinct from the updated LSTP results. The difference is
therefore not caused by including the exact
$\mathcal O(\Lam^{5/2})$ coefficient only in the LSTP construction.

Filter 3 is therefore refined to distinguish isolated denominator
roots from near-canceling pole-zero pairs. A pair is classified as a
Froissart doublet when both its normalized separation and normalized
residue are below $10^{-4}$, which the pair of
Eq.~\eqref{eq:doublet} satisfies by two orders of magnitude; varying
this threshold from $10^{-5}$ to $10^{-3}$ leaves all classifications
unchanged. Applying the refined filter to the LSTP scan changes no
candidate classification. The mapped roots that exclude the
$\Lambda_0=1$ curves have normalized separations $0.27$ and $0.71$,
independent of $\alpha$, and the closest pair in the scan, on the
$\Lambda_0=0.5$ curves, has separation and residue near
$1.1\times10^{-3}$; those curves also fail the bounds and monotonicity
conditions. The unique $\Lambda_0=4$ survivor of
Sec.~\ref{sec:results} is unchanged.

\section{Results}
\label{sec:results}

\subsection{From a broad band to a unique LSTP curve}

\begin{figure*}[!t]
  \centering
  \includegraphics[width=0.65\textwidth]{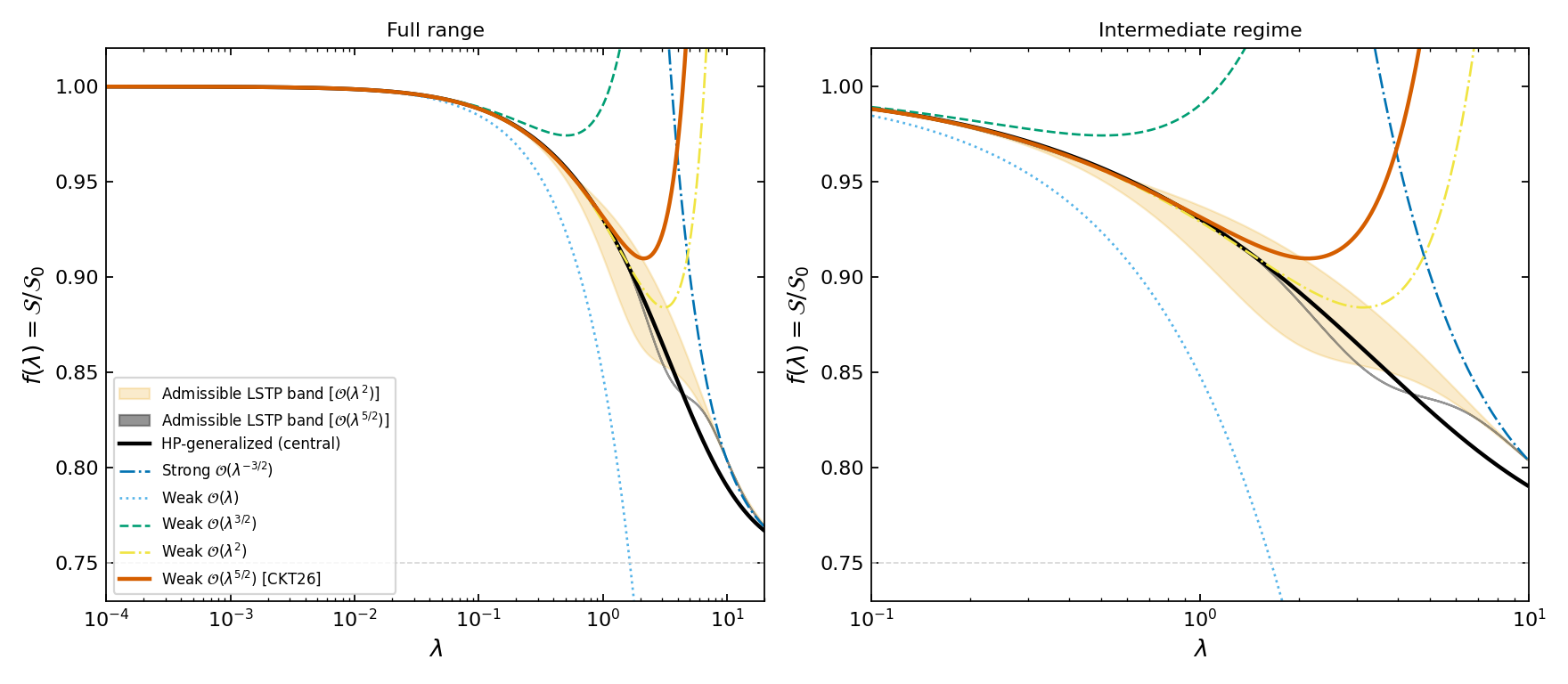}
  \caption{Comparison of the old $\mathcal O(\Lam^2)$ admissible LSTP band and
  the updated band obtained after imposing the exact $\mathcal O(\Lam^{5/2})$
  weak-side coefficient. The solid black line is the HP-generalized
  central curve defined in Ref.~\cite{Tantary:2026ensemble}, retained
  unchanged as a reference for comparing routes. Left: full range. Right:
  intermediate-coupling regime. The updated dark band collapses onto a
  single visible curve. The beige $\mathcal O(\Lam^2)$ band is the spread
  of admissible scan curves, not a statistical error band; a curve built
  with additional physical information is therefore not required to lie inside
  it. The updated survivor has $\Lambda_0=4$, for which no admissible
  curve existed in the $\mathcal O(\Lam^2)$ scan.}
  \label{fig:bands}
\end{figure*}

\begin{figure*}[!t]
  \centering
  \includegraphics[width=0.65\textwidth]{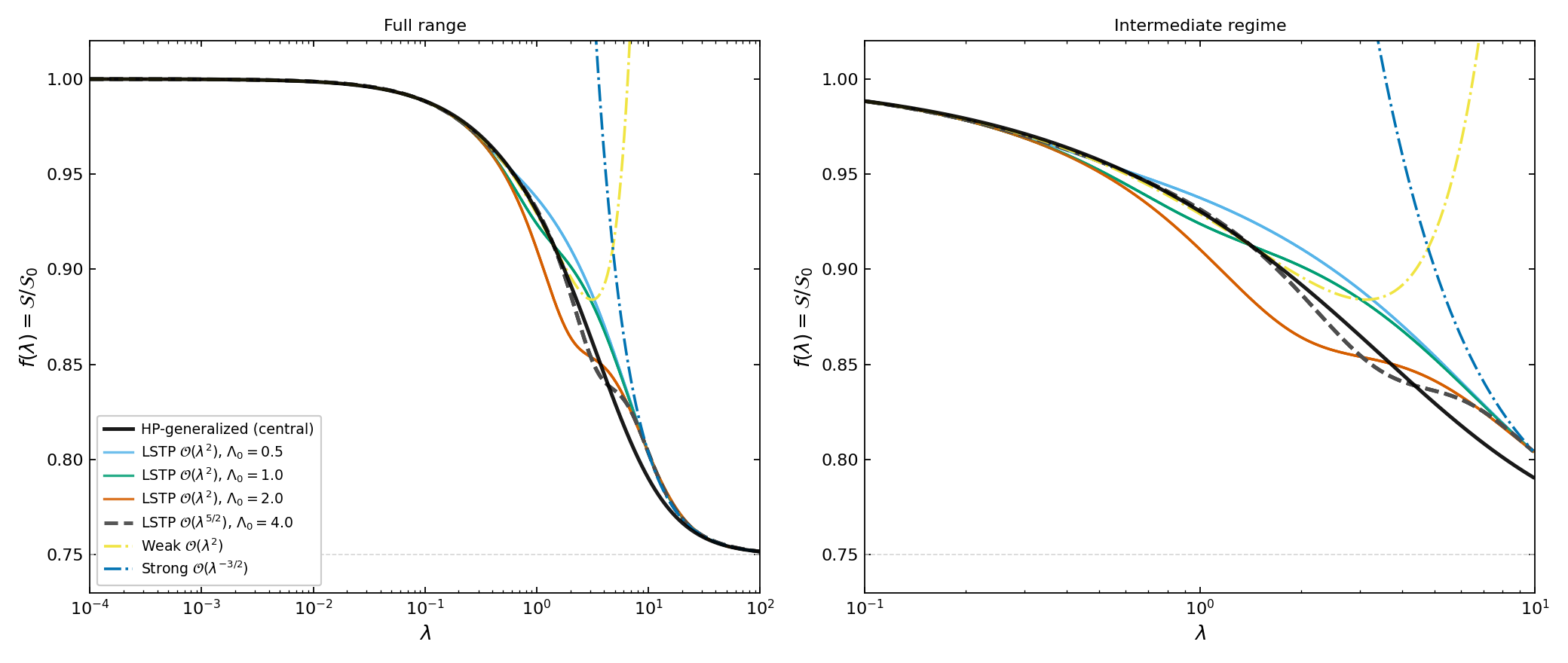}
  \caption{Individual LSTP survivor curves before and after imposing the exact
  $\mathcal O(\Lam^{5/2})$ coefficient. Since the three values of $\alpha$
  coincide at each $\Lambda_0$, the nine $\mathcal O(\Lam^2)$ survivors
  trace only three distinct curves (one per $\Lambda_0$, solid lines), and
  the three $\mathcal O(\Lam^{5/2})$ survivors collapse onto a single dashed
  curve at $\Lambda_0=4.0$. Left: full range. Right: intermediate-coupling
  zoom. The black line is the HP-generalized central curve
  of Ref.~\cite{Tantary:2026ensemble}, retained unchanged.}
  \label{fig:allcurves}
\end{figure*}

The change in the admissible set is summarized in Table~\ref{tab:comparison}.
With only $\mathcal O(\Lam^2)$ weak-side matching, the LSTP scan admits $9$
nominal survivors, but the three values of $\alpha$ give identical curves at
each surviving $\Lambda_0$, so only $3$ distinct curves remain
(Sec.~\ref{sec:method}).
Their crossover values span the wide interval reported in
Ref.~\cite{Tantary:2026ensemble},
\begin{equation}
\label{eq:oldrange}
\Lam_c^{\rm old}\in[2.9520,\,6.7321]\,.
\end{equation}

Once the exact $\mathcal O(\Lam^{5/2})$ coefficient is imposed, only
three of the twelve $(\alpha,\Lambda_0)$ points in the scan pass all
three admissibility filters:
\begin{equation}
(\alpha,\Lambda_0)=(0.5,4.0),\ (1.0,4.0),\ (2.0,4.0)\,.
\end{equation}
The remaining nine combinations are rejected as follows: the
$\Lambda_0=0.5$ curves fail the bounds and monotonicity conditions,
and the $\Lambda_0=2.0$ curves fail monotonicity. The
$\Lambda_0=1.0$ curves satisfy the bounds and monotonicity
conditions on the physical axis but are excluded by filter~3 because
of isolated denominator roots under the global mapped-root criterion.
Thus the uniqueness of the survivor is conditional on that criterion,
inherited from Ref.~\cite{Tantary:2026ensemble}. The doublet-aware refinement of
Sec.~\ref{sec:hp_update} leaves these classifications unchanged and is
applied consistently to both scans. The three surviving points share the same
$\Lambda_0=4.0$ and, because of the $\alpha$-degeneracy noted in
Sec.~\ref{sec:method}, give numerically identical curves on the full
evaluation grid. The updated admissible set is a single LSTP curve,
with crossover
\begin{equation}
\label{eq:newlc}
\Lam_c^{\rm new}\simeq 4.7863,
\qquad
f(\Lam_c^{\rm new})\simeq 0.8370886\,.
\end{equation}
This value lies inside the old admissible range
(Eq.~\eqref{eq:oldrange}), so the updated weak-side matching
produces a crossover value that was already within the range allowed
by the earlier scan. The
$\mathcal O(\Lam^2)$ column of Table~\ref{tab:comparison} reports an
interval because three distinct old curves have three different
crossover values; the $\mathcal O(\Lam^{5/2})$ column is a single
number because only one curve survives.

\begin{table}[t]
\caption{Before/after comparison of the admissible LSTP ensemble.
``Nominal survivors'' is the number of $(\alpha,\Lambda_0)$ points
that pass all three admissibility filters; ``distinct curves'' counts
how many of those are numerically different, since the three $\alpha$
values at fixed $\Lambda_0$ always coincide within the $\beta=0$
subfamily. Columns correspond to the $\mathcal O(\Lam^2)$ scan of
Ref.~\cite{Tantary:2026ensemble} and the exact
$\mathcal O(\Lam^{5/2})$ scan of this work.}
\label{tab:comparison}
\begin{ruledtabular}
\begin{tabular}{lcc}
Quantity & $\mathcal O(\Lam^2)$ & $\mathcal O(\Lam^{5/2})$ \\
\hline
Nominal survivors   & 9       & 3       \\
Distinct curves     & 3       & 1       \\
$\Lam_c$ range      & $[2.9520,\,6.7321]$  & $4.7863$  \\
Max band width ($1\!\le\!\Lam\!\le\!10$) & 0.047 & $<10^{-15}$ \\
\end{tabular}
\end{ruledtabular}
\end{table}

Figure~\ref{fig:bands} shows the old and updated admissible bands
together with the unchanged HP central curve. The new band is, within
numerical resolution, a single curve: the maximum pointwise width on
$1\le\Lam\le10$ drops from $0.047$ to below $10^{-15}$
(Table~\ref{tab:comparison}). The revised weak-coupling matching
changes the series values at the intermediate-coupling matching
points. The resulting curves with $\Lambda_0\in\{0.5,1.0,2.0\}$ fail
one or more of the filters listed above, while $\Lambda_0=4.0$ passes
all three. Thus the reduction of the band follows from the revised
matching, not from a change in the filter tolerances.
This is a statement about the twelve $(\alpha,\Lambda_0)$ candidates of
the stated scan. Because the updated scan changes both the
weak-coupling truncation and the weak-side matching points, it does
not isolate their separate effects.

Figure~\ref{fig:allcurves} shows the collapse curve by curve. The
left panel shows the full coupling range; the right panel zooms into
the intermediate regime where the spread is largest. In the
$\mathcal O(\Lam^2)$ scan, the nine nominal survivors show a visible
spread, while after imposing the exact $\mathcal O(\Lam^{5/2})$
coefficient all three survivors at $\Lambda_0=4.0$ are numerically
identical and appear as a single dashed curve.
\subsection{Strong-coupling residual estimator}
\label{sec:S3hat}

The known large-$N_c$ expansion~\eqref{eq:strong} is truncated at
$\mathcal O(\Lam^{-3/2})$; the coefficient of the next term
$\Lam^{-3}$ is not analytically known.
Following Ref.~\cite{Tantary:2026ensemble}, we define the
finite-coupling estimator
\begin{equation}
\label{eq:S3hat}
\hat S_3(\Lam_\ast) = \frac{4}{3}\Lam_\ast^{3}
\!\left[f(\Lam_\ast)
-\frac{3}{4}\!\left(1+\frac{15\zeta(3)}{8}\Lam_\ast^{-3/2}\right)\right].
\end{equation}
Here $\Lam_\ast$ is the coupling at which we evaluate the
interpolant. The idea is direct: if the full strong-coupling expansion
reads
$f=\frac{3}{4}[1+\frac{15\zeta(3)}{8}\Lam^{-3/2}+S_3\Lam^{-3}+\cdots]$,
then subtracting the two known terms and multiplying by
$(4/3)\Lam_\ast^3$ isolates $S_3$ as $\Lam_\ast\to\infty$. At finite
$\Lam_\ast$ the residual $f(\Lam_\ast)-(3/4)(1+(15\zeta(3)/8)\Lam_\ast^{-3/2})$
reflects both higher-order terms in the true strong expansion and the
rational structure of the interpolant at intermediate $\Lam$.
$\hat S_3(\Lam_\ast)$ is therefore a finite-coupling \emph{estimator}
of the interpolant's residual at $\Lam_\ast$, not a clean asymptotic
extraction of $S_3$ itself.

For any LSTP curve, $\hat S_3$ vanishes at the strong-side matching
points $\Lam\in\{10,30,100,300\}$ since the interpolant is forced to
match $f_{\rm strong}$ there by construction. We therefore evaluate at
$\Lam_\ast=20$, which lies between matching points and is not directly
constrained.

Before the weak-side upgrade, the admissible LSTP band contained
three distinct curves, so $\hat S_3^{\rm old}(20)$ was not a single
number but a range:
\begin{equation}
\hat S_3^{\rm old}(20)\in[9.09,\,13.03]\,.
\end{equation}
After imposing the exact $\mathcal O(\Lam^{5/2})$ coefficient, the
LSTP band reduces to a single curve (Sec.~\ref{sec:results}), so
$\hat S_3^{\rm LSTP}(20)$ becomes a single number. The unchanged HP
central curve gives a separate value:
\begin{align}
&\hat S_3^{\rm LSTP}(20) \approx +2.14\,,\label{eq:S3LSTP}\\
&\hat S_3^{\rm HP}(20)   \approx -21.86\,.\label{eq:S3HP}
\end{align}
The two routes disagree in both magnitude and sign: the LSTP route
gives a positive residual at $\Lam_\ast=20$, while the HP route gives a
strongly negative one. This comparison uses different weak-coupling
truncations: the HP value comes from the old $[5/5]$ curve, which does
not include $A_{5/2}$. It mirrors the crossover offset of Sec.~\ref{sec:crossover}:
the exact $\mathcal O(\Lam^{5/2})$ coefficient
pins $\hat S_3^{\rm LSTP}$ to a single value but cannot determine the
sign of the true asymptotic coefficient, which requires an independent
strong-coupling calculation.

The comparison in Eqs.~\eqref{eq:S3LSTP}--\eqref{eq:S3HP} is
asymmetric: the LSTP route uses the exact $\mathcal O(\Lam^{5/2})$
coefficient, while the $[5/5]$ HP curve is retained at
$\mathcal O(\Lam^2)$. The comparison using the same known weak- and
strong-coupling expansions is supplied by the
$[6/6]$ curve of Sec.~\ref{sec:hp_update}, evaluated away from its
localized doublet, which gives
$\hat S_3^{\rm HP66}(20)\simeq-29.77$~[Eq.~\eqref{eq:hp66results}], on
the same side as the $[5/5]$ value and farther from the LSTP result.
The LSTP--HP difference therefore persists when both routes carry the
exact coefficient. Neither route is preferred a priori; no
independent finite-$\Lam$ benchmark is currently available to decide
between them. We therefore treat
Eqs.~\eqref{eq:S3LSTP}--\eqref{eq:S3HP} as a construction-dependent
comparison, with $\hat S_3^{\rm HP}$ a reference value from
the old $[5/5]$ curve, to be revisited once the next strong-coupling
coefficient is computed.

\subsection{Crossover collapse and route dependence}
\label{sec:crossover}

\begin{figure}[t]
  \centering
  \includegraphics[width=0.95\columnwidth]{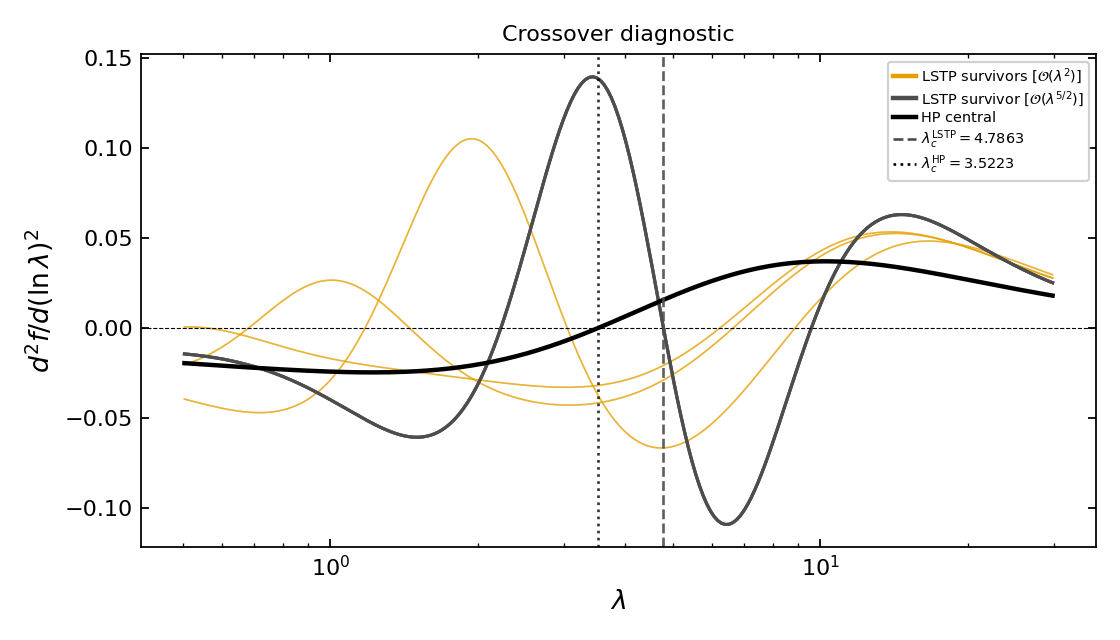}
  \caption{Curvature diagnostic $d^2 f/d(\log\Lam)^2$ for the old and
  updated LSTP survivors and the retained HP central curve. Thin
  orange curves: the nine $\mathcal O(\Lam^2)$ LSTP survivors of
  Ref.~\cite{Tantary:2026ensemble}; because the three $\alpha$ values
  give identical curves at fixed $\Lambda_0$ (Sec.~\ref{sec:method}),
  these collapse into only three visually distinct curves, one per
  surviving $\Lambda_0$ in the old scan. Thick dark-gray curve: the unique
  $\mathcal O(\Lam^{5/2})$ LSTP survivor. Thick solid black curve: the
  HP-generalized central curve from Ref.~\cite{Tantary:2026ensemble},
  retained unchanged. Vertical lines mark the LSTP crossover
  $\Lam_c^{\rm LSTP}=4.7863$ (dashed gray) and the HP
  crossover $\Lam_c^{\rm HP}=3.5223$ (dotted black). The old survivors
  span a broad inflection range, while the updated LSTP family collapses
  to a unique crossing. The large-amplitude oscillations in the new LSTP
  curvature are a feature of second-derivative diagnostics:
  $f(\Lam)$ itself remains monotone and within bounds on the full grid.}
  \label{fig:crossover}
\end{figure}

The curvature diagnostic provides an independent check. In the old
ensemble, the admissible LSTP survivors had inflection points spread
over the wide interval in Eq.~\eqref{eq:oldrange}. In the updated
ensemble the three $(\alpha,\Lambda_0=4.0)$ survivors are numerically
identical (Sec.~\ref{sec:results}), so they trace a single curvature
curve with a single inflection crossing at the
$\Lam_c^{\rm new}$ of Eq.~\eqref{eq:newlc}. In
Fig.~\ref{fig:crossover} one therefore sees only one LSTP
curve in the updated layer; the three survivors coincide.

The updated LSTP curve does not merge with the HP central
curve. The crossover values from LSTP, HP, and the M\"uller Pad\'e are
\begin{equation}
\Lam_c^{\rm LSTP}\simeq 4.79,\quad
\Lam_c^{\rm HP}\simeq 3.52,\quad
\Lam_c^{\rm M}\simeq 3.14\,,
\end{equation}
where $\Lam_c^{\rm M}$ is the single-curve Pad\'e value of
Ref.~\cite{Muller:2025}. The HP and M\"uller values are close to each other, while the updated
LSTP crossover is noticeably higher. Three observations put this
offset in context.

\emph{First}, $\Lam_c^{\rm LSTP}$ falls inside the old admissible
range $[2.95,\,6.73]$ of Ref.~\cite{Tantary:2026ensemble}, so the
updated scan selects a crossover value that the earlier scan
already allowed. The updated scan narrows the range; it does not
push the result outside it.

\emph{Second}, the comparison above is asymmetric by construction: the
LSTP route is the only one updated with the exact
$\mathcal O(\Lam^{5/2})$ coefficient. The HP ansatz and the M\"uller
Pad\'e were both built with weak-side matching through
$\mathcal O(\Lam^2)$ only. The test using the same known expansions is carried out in
Sec.~\ref{sec:hp_update}: the $[6/6]$ curve, evaluated away from its
doublet, gives
$\Lam_c^{\rm HP66}\simeq2.98$, below the $[5/5]$ value and farther
from the updated LSTP crossover. The upward shift of the LSTP
crossover is therefore not an artifact of the weak-side truncations
differing.

\emph{Third}, the updated LSTP scan changes both the weak-coupling
truncation and the weak-side matching points. The shift of
$\Lam_c$ relative to the HP and M\"uller curves is therefore a
property of the updated LSTP construction and cannot be attributed
to $A_{5/2}$ alone.

The updated scan therefore resolves the ambiguity
\emph{within} the LSTP route but not the difference between LSTP
and HP.

\section{Discussion}
\label{sec:discussion}

The updated scan changes the interpretation of constrained Pad\'e ensembles in
thermal $\SYM$. In the earlier $\mathcal O(\Lam^2)$ paper, the admissible
ensemble quantified the spread among the stated candidates. Here both the
weak-coupling truncation and the weak-side matching points are updated.
Within the same discrete scan, the revised matching reduces the old
three-curve spread to one distinct curve. This comparison tests the
combined matching prescription rather than the effect of $A_{5/2}$ alone.

Two points follow. First, the discrete LSTP scan is sensitive to the
weak-coupling matching prescription. Second, the persistence of the
HP/LSTP separation means that the presently known weak-coupling
coefficients are not enough to remove all interpolation ambiguity.
In the present problem, the relevant unknown is the next
strong-coupling coefficient, at $\mathcal O(\Lam^{-3})$.

The finite-coupling estimator $\hat S_3(\Lam_\ast)$ defined in
Eq.~\eqref{eq:S3hat} is no longer a broad-band diagnostic within the LSTP
route.
The updated matching prescription reduces the old LSTP range
$\hat S_3^{\rm old}(20)\in[9.09,13.03]$ to the unique value
$\hat S_3^{\rm LSTP}(20)\approx+2.14$~[Eq.~\eqref{eq:S3LSTP}], while the
HP route gives $\hat S_3^{\rm HP}(20)\approx-21.86$~[Eq.~\eqref{eq:S3HP}].
The sign reversal between the two routes is a concrete finite-$\Lam$
diagnostic. The HP
entry is a reference value from the old $[5/5]$ curve, which
does not carry $A_{5/2}$, so we read the difference as
construction-dependent rather than as a statement about the true sign
of the asymptotic coefficient $S_3$. With both constructions using the
same known expansions, the
$[6/6]$ curve gives $\hat S_3^{\rm HP66}(20)\simeq-29.8$
(Sec.~\ref{sec:hp_update}), on the same side as the $[5/5]$ value, so
the difference between the routes persists. Resolving the true
asymptotic sign of $S_3$ still requires the strong-coupling expansion
beyond $\mathcal O(\Lam^{-3/2})$.

Rather than asking only whether an ansatz can ``predict'' the next coefficient,
one can ask which interpolation families survive once the next coefficient is
known exactly, and how much of the remaining spread is route dependence rather
than scan uncertainty. The updated scan turns
the old ensemble from a prediction problem into a selection problem.

The unique updated LSTP curve can now be used to revisit strong-side
estimators such as $S_3$ under tighter weak-side constraints. The route
difference that remains is intrinsic to the present constructions
(Sec.~\ref{sec:hp_update}), not an artifact of the weak-side truncations differing.

\section{Conclusions}
\label{sec:conclusions}

We have revisited constrained Pad\'e interpolation in thermal $\SYM$ after the
weak-coupling expansion was upgraded from $\mathcal O(\Lam^2)$ to exact
$\mathcal O(\Lam^{5/2})$. Using the same LSTP construction and admissibility
filters as in the earlier ensemble paper, we find:
\begin{enumerate}
\item the old $9$ nominal LSTP survivors ($3$ distinct curves) collapse to $3$
nominal survivors that are numerically identical, i.e.\ to a single distinct
LSTP curve within the stated scan, with pointwise band width falling to zero
within numerical resolution;
\item the old crossover range
$\Lam_c\in[2.9520,\,6.7321]$ collapses to the unique value
$\Lam_c=4.7863$, with $f(\Lam_c)=0.8371$. This value lies inside the
old range. Because both the weak-coupling truncation and the matching
points change in the updated scan, its shift relative to the HP
($3.52$) and M\"uller ($3.14$) values cannot be attributed to
$A_{5/2}$ alone;
\item the HP central curve remains distinct at $\Lam_c=3.5223$. The
comparison is asymmetric: only the LSTP route is updated with
$A_{5/2}$. The updated scan therefore removes the spread among the
stated LSTP candidates but not the difference between the two routes;
\item the minimal HP$[6/6]$ update including $A_{5/2}$ is uniquely
determined and develops a near-canceling pole-zero pair at
$\Lam\simeq0.17956$ ($\Lam_D=0.179556755567\ldots$,
$\Lam_N=0.179556946766\ldots$), with normalized separation and residue
both of order $10^{-6}$: a spurious Froissart doublet, not a physical
singularity. The exactly matched curve is retained and evaluated away
from the pair; dividing the pair out, as a stability check, changes the
results by less than $10^{-4}$.
The curve gives $\Lam_c^{\rm HP66}\simeq2.98$ and
$\hat S_3^{\rm HP66}(20)\simeq-29.8$;
\item the strong-coupling residual estimator $\hat S_3(20)$ differs
between the updated LSTP curve ($+2.14$), the old HP$[5/5]$ reference
($-21.86$, without $A_{5/2}$), and the HP$[6/6]$ curve carrying $A_{5/2}$
($-29.8$). The LSTP--HP difference persists when both routes carry the
exact coefficient; its interpretation as the true $S_3$ requires the
next strong-coupling coefficient.
\end{enumerate}

The main result is not simply that the band narrows: the updated
weak-side matching selects a unique LSTP interpolant.
The next strong-coupling coefficient, at $\mathcal O(\Lam^{-3})$,
remains unknown. Obtaining it would require higher-order
$\alpha'$ corrections to the type-IIB string effective action beyond
those entering the known $\mathcal O(\Lam^{-3/2})$ result, and is
therefore a major calculation in itself.

\begin{acknowledgments}
The work of Q.D.\ is supported by the
Guangxi Natural Science Foundation under Grant No.~2023GXNSFBA026027. U.T.\ acknowledges support
from the Department of Mathematics and Natural Sciences at Prince
Mohammad Bin Fahd University. We thank Michael Strickland for earlier collaborations on
Pad\'e interpolation in thermal $\mathcal N{=}4$ SYM, and thank Margaret
E.\ Carrington and Gabor Kunstatter for collaboration with U.T. on the $\mathcal
O(\Lam^{5/2})$ thermodynamics calculation.
\end{acknowledgments}

\end{document}